\documentstyle[12pt]{article}
\begin{document}
\rightline{TUHE-9661}
\rightline{hep-ph/9606326}
\begin{center}
{\bf Mixing of $\Xi_c-\Xi_c^{\prime}$ baryons}\\
\vspace{16pt}
Jerrold Franklin\\
{\it Department of Physics,Temple University,\\
Philadelphia, Pennsylvania 19122-8062}\\
June, 1996
\end{center}
\begin{abstract}

The mixing angle between the $\Xi_c$ and $\Xi_c^{\prime}$ baryons
is shown to be small, with a negligible shift in the $\Xi_c$ masses.
\end{abstract}
\vspace{.5in}

  Some concern has been expressed\cite{lebed,ito} that mixing between the
$\Xi_c$ and $\Xi_c^{\prime}$ baryons could have a significant effect on their
masses. However,
we show here that this mixing is small and shifts the $\Xi_c$ and
$\Xi_c^{\prime}$ masses by a negligible amount.\cite{cutk}

The theory for mixing between  ``flavor-degenerate" baryons which, like the
$\Xi_c$ and
$\Xi_c^{\prime}$, contain three different flavor quarks that  differ only in
their internal spin
composition has been developed in Ref. \cite{fl}.
The $\Xi_c$ and $\Xi_c^{\prime}$ are each composed of three different
flavored quarks, n, s, and c.
(We use the quark symbol n to refer to either the u or d quark.)
The n and s quarks are in a
relative spin 0 state for the $\Xi_c$, and in a spin 1 state for the
$\Xi_c^{\prime}$.
The physical $\Xi_c$ and $\Xi_c^{\prime}$ states with masses M and M$^{\prime}$
can be
written in terms of unmixed quark model states
$|nsc>$ and $|nsc^{\prime}>$ as
\begin{equation}
\Xi  =  +\cos\theta |nsc>+\sin\theta |nsc^{\prime}>,
\end{equation}
\begin{equation}
\Xi^{\prime}  =  -\sin\theta |nsc>+\cos\theta |nsc^{\prime}>.
\end{equation}
There will be a shift in mass from the pure quark state masses M$_{\circ}$ and
M$_{\circ}^{\prime}$
given by\cite{trig}
\begin{equation}
\delta M^{\prime} = -\delta M = \Delta\sin^2\theta,
\end{equation}
where $\Delta$=M$^{\prime}$-M is the mass difference between the two physical
states.

     If the symmetry breaking quark$_i$-quark$_j$ interaction is flavor
independent, except
for a dependence $\sim 1/m_im_j$, then it is shown in Ref. \cite{fl} that
\begin{equation}
\tan2\theta=\frac{\sqrt{3}(m_2-m_1)}{2m_3-m_1-m_2}.
\end{equation}
For $\Xi_c$ and $\Xi_c^{\prime}$ mixing, we use quark masses
\begin{equation}
m_1=m_n=330 MeV,\hspace{.2in} m_2=m_s=510 MeV,\hspace{.2in} m_3=m_c=1.6 GeV,
\end{equation}
with the result
\begin{equation}
\theta_{\Xi_c-\Xi_c^{\prime}} =3.8^{\circ},\hspace{.5in}
\delta M^{\prime}=0.4 MeV.
\end{equation}
This is our main result.  The small value for $\delta M^{\prime}$ depends on the
small ratio (m$_s$-m$_n$)/m$_c$=0.11, which is about the same for all quark
models.

     Although the assumption that the symmetry breaking quark$_i$-quark$_j$
interaction is
flavor independent except for a
 $\sim 1/m_im_j$ dependence is part of most baryon mass calculations, the
$\Xi_c$-$\Xi_c^{\prime}$ mixing can be estimated more generally without this
assumption.  In
fact for $\Sigma-\Lambda$ mixing in the light baryon sector, this is necessary
because the
magnetic interaction between quarks is not flavor independent.  For this case,
the
$\Sigma-\Lambda$ mixing angle can be written in terms of light baryon masses
as\cite{trig}
\begin{equation}
\sin2\theta_{\Sigma-\Lambda}=\frac{(\Sigma^{-}-\Sigma^{+})
-(\Sigma^{*-}-\Sigma^{*+})}{\sqrt{3}(\Sigma^0-\Lambda)},
\end{equation}
where the baryon symbol stands for its mass.
For $\Sigma-\Lambda$ mixing this leads to\cite{eqs}
\begin{equation}
\theta_{\Sigma-\Lambda} =0.8^{\circ}\pm .2^{\circ},\hspace{.5in}
\delta M^{\prime}=0.015 MeV.
\end{equation}
Although the mass shift is negligible, the  $\Sigma-\Lambda$ mixing does lead to
a significant shift in the $\Lambda$ magnetic moment
of -0.045$\pm$.002 nuclear magnetons which should be included in any quark model
calculation of the $\Lambda$ magnetic moment.

Equation (7) for $\Sigma-\Lambda$ mixing can be extended to the charmed baryons
by simply replacing the uds quarks
of the $\Sigma^0$ and $\Lambda$ in Eq. (7) by the nsc quarks of the
$\Xi_c$ and $\Xi_c^{\prime}$.  That is, as in Ref. \cite{cb1}, we make the quark
substitutions
\begin{equation}
u\rightarrow u,\hspace{.2in}d\rightarrow s,\hspace{.2in}s\rightarrow c
\end{equation}
everywhere in Eq. (7).  This results in
\begin{equation}
\sin2\theta_{\Xi_c-\Xi_c^{\prime}}=\frac{(\Omega_c^{0}-\Sigma_c^{++})
-(\Omega_c^{*0}-\Sigma_c^{*++})}{\sqrt{3}(\Xi_c^{\prime +}-\Xi_c^+)}.
\end{equation}

The $\Omega_c^*$ baryon has not been observed yet so we use the sum
rule\cite{applic}
\begin{equation}
\Omega_c^{*0}-\Omega_c^0=2(\Xi_c^{*+}-\Xi_c^{\prime +}+\delta M^{\prime})
-(\Sigma_c^{*++}-\Sigma_c^{++})
\end{equation}
to replace Eq. (9) by
\begin{equation}
\sin2\theta_{\Xi_c-\Xi_c^{\prime}}=\frac{2[(\Sigma_c^{*++}-\Sigma_c^{++})
-(\Xi_c^{*+}-\Xi_c^{\prime +})]}
{(\Xi_c^{\prime +}-\Xi_c^+)(\sqrt{3}+\tan\theta_{\Xi_c-\Xi_c^{\prime}})}.
\end{equation}
A similar equation holds for the $\Xi_c^0$ and $\Xi_c^{\prime 0}$
baryons if the quark substitution $u\rightarrow d$ is made in Eq. (12).
The tan$\theta_{\Xi_c-\Xi_c^{\prime}}$ term in Eq. (12) is a small  correction
that permits the use of only physical baryon masses
in the equation.

We use the masses (in MeV)
\begin{equation}
\hspace{.1in}\Xi_c^{*+}=2644.6\pm2.3\cite{cleo96-4},\hspace{.2in}
\Xi_c^{\prime +}=2563\pm15\cite{wa},\hspace{.2in}
\Xi_c^{+}=2465.1\pm1.6\cite{pdg},
\end{equation}
\begin{equation}
\Sigma_c^{*++}=2530\pm7\cite{skat},\hspace{.4in}
\Sigma_c^{++}=2453\cite{pdg}
\end{equation}
to get the results
\begin{equation}
\theta_{\Xi_c-\Xi_c^{\prime}} =-2^{\circ}\pm{6^{\circ}},
\hspace{.5in}\delta M^{\prime}<2 MeV.
\end{equation}
This is consistent with Eq. (6), but better experimental accuracy would be
required to make a significant comparison.  We have used the 1 $\sigma$
experimental error to get the 2 MeV limit on $\delta M^{\prime}$.  This is
already negligible for most applications,
and would be expected to decrease as the experimental accuracy improves.

 Our conclusion is that $\Xi_c-\Xi_c^{\prime}$ mixing can be consistently
neglected in quark model calculations.  A similar conclusion holds for b quark
baryons where the quark substitution
c$\rightarrow$b would lead to even smaller $\Xi_b-\Xi_b^{\prime}$ mixing.


\begin{thebibliography}{8}
\bibitem{trig}Reference[4] used small angle approximations, but the
trigonometry is treated exactly here.  Also note that Ref. [4] used a
different convention for determining the primed state.
\bibitem{eqs}The $\Sigma-\Lambda$ mixing parameters given here come from
using Eqs. (48)-(51) of Ref. [4].  We use newer experimental masses so that the
numbers given here are different than those in Ref.[4].
\end{thebibliography}
\end{document}